\documentclass[aps,pra,reprint,superscriptaddress,longbibliography,floatfix]{revtex4-2}

\usepackage{amsmath,amssymb,bm}
\usepackage{booktabs}
\usepackage{graphicx}
\usepackage[hidelinks]{hyperref}
\usepackage[version=4]{mhchem}
\usepackage{siunitx}

\graphicspath{{figures/}}
\newcommand{\rvec}{\bm{r}}
\newcommand{\Bvec}{\bm{B}}
\newcommand{\Avec}{\bm{A}}
\newcommand{\iu}{\mathrm{i}}
\newcommand{\eu}{\mathrm{e}}

\newcommand{\au}{\,\mathrm{a.u.}}
\newcommand{\cmone}{\mathrm{cm}^{-1}}

\begin{document}

\title{Gauge-including neural-network quantum Monte Carlo for molecules in magnetic fields}

\author{Chengye L\"u}
\thanks{These authors contributed equally to this work.}
\affiliation{Key Laboratory of Computational Physical Sciences (Ministry of Education), Institute of Computational Physical Sciences, State Key Laboratory of Surface Physics, and Department of Physics, Fudan University, Shanghai 200433, China}
\author{Weizhong Fu}
\thanks{These authors contributed equally to this work.}
\affiliation{Key Laboratory of Computational Physical Sciences (Ministry of Education), Institute of Computational Physical Sciences, State Key Laboratory of Surface Physics, and Department of Physics, Fudan University, Shanghai 200433, China}
\author{Xin-gao Gong}
\affiliation{Key Laboratory of Computational Physical Sciences (Ministry of Education), Institute of Computational Physical Sciences, State Key Laboratory of Surface Physics, and Department of Physics, Fudan University, Shanghai 200433, China}
\author{Hongjun Xiang}
\email{hxiang@fudan.edu.cn}
\affiliation{Key Laboratory of Computational Physical Sciences (Ministry of Education), Institute of Computational Physical Sciences, State Key Laboratory of Surface Physics, and Department of Physics, Fudan University, Shanghai 200433, China}

\begin{abstract}
External magnetic fields, through their coupling to orbital and spin motion, complicate 
the correlated electronic states  and impose coordinate-dependent phases on the wavefunction, 
thereby making accurate electronic structure calculations substantially more demanding.
Recently, neural network-based quantum Monte Carlo (NNQMC) has emerged as a highly accurate 
approach to study nucleus-free systems in magnetic fields. For molecular systems, however, 
things get more complicated as the magnetic field would introduce a rapidly varying phase 
in the region far from the gauge origin. Here we introduce a gauge-including phase factor 
that acts directly on the  full many-electron wavefunction and accounts for the prescribed magnetic phase, 
leaving a smoother correlated residual for the network to learn.
This factor greatly improves molecular translation consistency and size consistency, 
providing a route for studying systems in magnetic fields with NNQMC.
Upon this approach, we reproduce weak-field magnetizabilities and strong-field bond contraction in \ce{H2}. 
We further apply the method to selected transitions in the \ce{CN} red and \ce{C2} Swan systems 
at magnetic fields relevant to white dwarfs. The \ce{CN} transition exhibits a much larger field-induced 
shift than its \ce{C2} counterpart, suggesting its potential as a probe of white-dwarf magnetic fields.
\end{abstract}

\maketitle

\section{Introduction}

The electronic structure of molecules in strong magnetic fields connects fundamental
chemical bonding with astrophysical
spectroscopy~\cite{ferrario2015,lai2001,harding2006}. When $B\sim1\au$
($2.35\times10^5\,\mathrm{T}$), magnetic and Coulomb interactions become comparable, and
the field is no longer a small perturbation. Diamagnetic contraction, orbital
paramagnetic response, and field-induced bonding mechanisms can then appear, including
paramagnetic bonding in triplet molecules under strong perpendicular
fields~\cite{lange2012}. Fields of this magnitude occur in magnetic white-dwarf
atmospheres, while still stronger fields characterize neutron-star
environments~\cite{ferrario2015,lai2001,harding2006}. To diagnose physical conditions,
quantitative many-electron calculations are needed for strong-field chemistry and
astronomical modeling.

Gauge-including atomic orbitals (GIAOs), also known as London orbitals, were
consequently developed to attach a prescribed field-dependent phase to each one-electron
basis function and thereby avoid the gauge-origin
dependence~\cite{london,ditchfield1974,ruud1993,tellgren2008,lange2012,
stopkowicz2015,zalialiutdinov2025}.
With GIAOs, traditional quantum chemistry methods can provide important benchmarks
~\cite{stopkowicz2015,lehtola2020diatomic,zalialiutdinov2025}, but the many-body
difficulty remains. For example, single-reference coupled-cluster methods can become
unreliable when fields induce level crossings, or
near-degeneracies~\cite{bartlett2007,stopkowicz2015}. Multireference methods offer
alternatives, but each introduces its own limitation. Full configuration interaction
(FCI) is exact within a finite one-electron basis but is limited by the exponential
growth of its determinant space~\cite{booth2013}. Multiconfigurational self-consistent
field (MCSCF) avoids this full-space cost through a selected active
space~\cite{ruud1995}, whose choice requires substantial system-specific
expertise~\cite{stein2016}. These methodological difficulties are further compounded by
a basis-set limitation: standard Gaussian basis sets, optimized for near-spherical
zero-field atoms, can poorly represent the strongly anisotropic orbitals induced by
strong fields~\cite{lehtola2020diatomic}.

In recent years, neural network-based quantum Monte Carlo (NNQMC) has emerged as a
complementary approach that optimizes a correlated many-electron wavefunction
parameterized by a neural network rather than a fixed basis set or active-space
truncation~\cite{ferminet,deepqmc,psiformer,nndmc}. This is particularly attractive in
magnetic fields, where orbital anisotropy, near-degeneracies, and correlation effects
can change simultaneously and are difficult to anticipate in a compact orbital
expansion. Remarkably, NNQMC has proven to be highly powerful not only for eigenstate
calculations of zero-field molecules and
solids~\cite{psiformer,nndmc,fu2024variance,pfau2024accurate,li2024spin,li2022ab,
li2024computational,fu2026empowering,scherbela2025accurate},
but also for accurately characterizing their static and even dynamic responses to
external electric fields~\cite{li2024polarization,fu2026towards}, showing its great
potential to study systems in magnetic fields. Recently, NNQMC has already been applied
to magnetic-field problems without molecular nuclei, including fractional quantum Hall
states of electron gases~\cite{qian2025deephall,teng2025electron_gas,abouelkomsan2026fqh}. Molecular systems
introduce a distinct challenge, that is, a rigid translation of the nuclei and electrons
relative to the gauge origin produces a rapidly varying magnetic-translation phase in
the many-electron wavefunction. At the orbital level, GIAOs incorporate the
corresponding prescribed field-dependent phase, but their one-electron construction
cannot be applied directly to NNQMC. Without an analogous many-electron factor, the
network must learn magnetic phase together with the correlated part of the wavefunction.

In this work, we introduce a gauge-including phase factor that acts directly on the full
many-electron wavefunction and analytically supplies the prescribed magnetic phase,
allowing the network to focus on learning many-electron correlations. Rigid-translation
tests and phase diagnostics demonstrate improved translational consistency and a
smoother learned residual. Also, energetic size consistency is greatly improved upon
incorporating this factor. Symmetry-restricted calculations further track matched
electronic shifts in the \ce{CN} red and \ce{C2} Swan systems. Together, these results
establish a practical NNQMC framework for molecular electronic structure in finite
magnetic fields.

\begin{figure*}[t]
  \centering
  \includegraphics[width=1.0\textwidth]{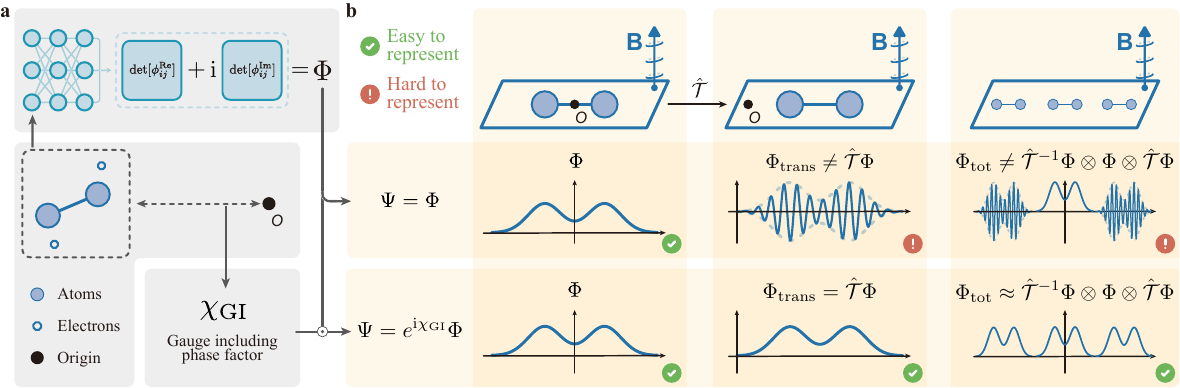}
  \caption[Gauge-including ansatz for finite-field NNQMC]{\textbf{Gauge-including ansatz
  for finite-field NNQMC.} \textbf{a,} The complex neural many-electron wavefunction $\Phi$ is
  multiplied by the prescribed phase factor to form $\Psi=\exp(\iu\chi_{\rm GI})\Phi$.
  \textbf{b,} The prescribed phase factor ensures magnetic-translation covariance
  and improves size consistency. Without the prescribed phase, the learned state must
  represent rapid phase structure under rigid translation and separated-fragment
  composition.  With the prescribed phase, these known phase contributions are supplied
  analytically, leaving a smoother residual, which is easier to represent by
  the neural network.}
  \label{fig:workflow}
\end{figure*}

\section{Results}

\subsection{Gauge-including real-space NNQMC in finite magnetic fields}

Figure~\ref{fig:workflow}a summarizes the proposed ansatz: a complex NNQMC wavefunction 
multiplied by an analytic phase factor. Figure~\ref{fig:workflow}b illustrates why this 
decomposition is useful for rigid translations and separated fragments in magnetic fields. 
Away from the gauge origin, the finite-field molecular wavefunction develops rapid spatial 
phase oscillations that are difficult for a neural network to represent accurately. 
Our formulation separates this rapidly varying phase from the smoother residual component, 
thereby reducing the representational burden on the neural network.

We implement the corresponding VMC
calculation with a complex-valued FermiNet-type antisymmetric wavefunction and a
magnetic local-energy estimator.  The trial state and proposed phase factor are
\begin{subequations}
\label{eq:main-gauge-including-state}
\begin{align}
  \Psi_{\theta}(\bm r)
  &=\eu^{\iu\chi_{\rm GI}(\bm r)}
  \Phi_{\theta}(\bm r),
  \label{eq:main-trial-state}\\
  \chi_{\rm GI}(\bm r)
  &=-\frac12\sum_i\Bvec\cdot
  \bigl[\bar{\bm R}_i\times (\rvec_i-\bm C_Z)].
  \label{eq:main-prescribed-phase}
\end{align}
\end{subequations}
Here $\Bvec$ is the external magnetic field, $\rvec_i$ is the coordinate of electron
$i$, and $\bm r=(\rvec_1,\ldots,\rvec_N)$ is the full electronic configuration of $N$
electrons; $\theta$ denotes the variational parameters, and $\Phi_{\theta}$
denotes the conventional complex FermiNet wavefunction. The smooth electron-dependent
nuclear coordinate $\bar{\bm R}_i$ for electron $i$ and the charge center $\bm C_Z$ are
\begin{equation}
\begin{aligned}
  \bar{\bm R}_i&=\sum_I w_{iI}\bm R_I,\qquad
  \sum_Iw_{iI}=1,\\
  \bm C_Z&=\frac{\sum_I Z_I\bm R_I}{\sum_I Z_I}.
 \end{aligned}
\label{eq:main-soft-center}
\end{equation}
Here $\bm R_I$ and $Z_I$ are the position and nuclear charge of nucleus $I$,
respectively, and softmax-type weights $w_{iI}$ define the nuclear anchor associated with
electron $i$. The proposed phase $\chi_{\rm GI}$ is constructed to encode the
coordinate-dependent transformation required for magnetic translations and gauge-origin
changes. Detailed definitions and derivations are presented in the Supplementary
Information (SI)~\cite{suppinfo}.

\subsection{Validation}

\begin{figure*}[t]
  \centering
  \includegraphics[width=0.98\textwidth]{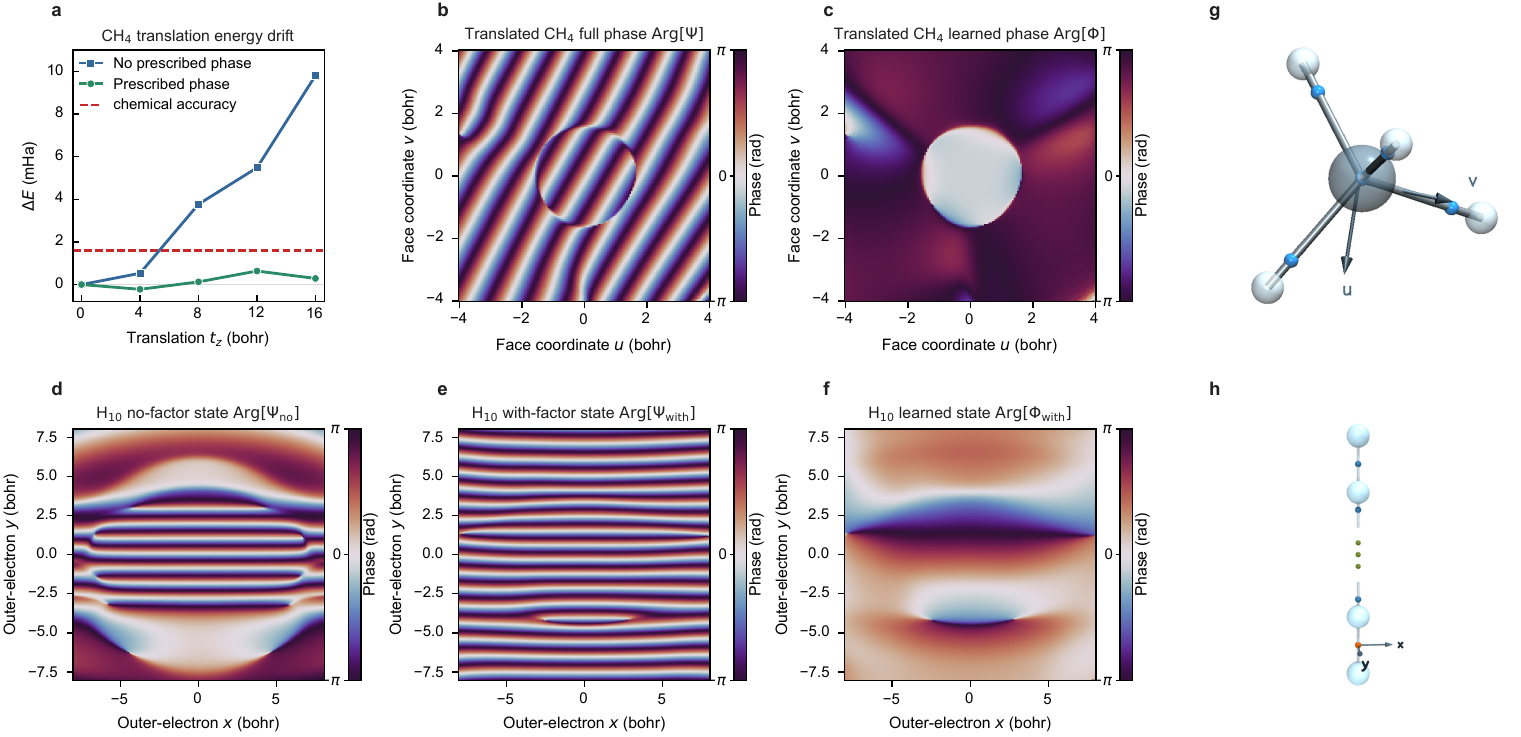}
  \caption[Magnetic translation consistency and phase structure in finite-field NNQMC]
  {\textbf{Magnetic translation consistency and phase structure in finite-field NNQMC.}
  \textbf{a,} Energy drift under rigid translation of \ce{CH4} at fixed gauge origin.
  \textbf{b,c,} Conditional one-electron phase slices of \ce{CH4}: in \textbf{b} the
  displayed phase is that of the full gauge-including state
  $\Psi=\exp(\iu\chi_{\rm GI})\Phi$, whereas \textbf{c} shows the learned neural state
  $\Phi$.  One electron is varied over the displayed two-dimensional slice plane that is 
  parametrized by the in-plane coordinates $u$ and $v$ and parallel to a tetrahedral face. All remaining electronic 
  coordinates are fixed at the recorded molecular anchor.  \textbf{d--f,}
  Conditional one-electron phase slices of \ce{H10}: \textbf{d} shows the full
  no-prescribed-phase state $\Psi_{\rm no}$, \textbf{e} the full prescribed-phase
  state $\Psi_{\rm with}=\exp(\iu\chi_{\rm GI})\Phi_{\rm with}$, and \textbf{f}
  the learned factor $\Phi_{\rm with}$.  One electron is varied over the
  transverse plane normal to the molecular chain, parametrized by the in-plane Cartesian
  coordinates $x$ and $y$, with its chain coordinate and all remaining electronic
  coordinates fixed at the recorded symmetric-chain anchor. \textbf{g,h,} Molecular structure diagrams showing the corresponding
  CH$_4$ and H$_{10}$ geometries and the scanned-electron directions.~\cite{liu2018visualization}}
  \label{fig:factor}
\end{figure*}

To assess the proposed factor, we first compare calculations with and without it. In the
\ce{CH4} translation test, displacing the molecule by $t_z=16$ bohr changes the energy
by $9.81$ mHa without the factor but by only $0.29$ mHa with it
(Fig.~\ref{fig:factor}a), corresponding to a roughly 34-fold reduction in the translation drift.
The residual drift of $0.1$ mHa level is consistent with the stochastic fluctuation of NNQMC.

The factor also improves the representation of the magnetic phase at fixed
molecular geometry. In the no-factor \ce{H10} slice, the neural state struggles to reproduce
the extended, rapidly oscillatory phase structure away from the gauge origin
(Fig.~\ref{fig:factor}d).  Including the prescribed phase captures this
long-range structure in the full wavefunction (Fig.~\ref{fig:factor}e), leaving a smoother residual 
for the network to represent (Fig.~\ref{fig:factor}f), as also seen in the \ce{CH4} maps
(Fig.~\ref{fig:factor}b,c).  For the displayed \ce{H10} configuration at bond length $R=3$ bohr, the
with-factor wavefunction also gives an energy $9.01$ mHa lower, 
indicating an improved variational representation. Consistently, magnetizability ablations for 
\ce{CH4} reported in the SI favor the with-factor formulation~\cite{suppinfo}.

Energetic size consistency is tested independently at $R=10$ bohr and $B=(1,0,0)\au$ by
comparing matched \ce{H}--\ce{H8} calculations using each phase arm's single-atom
\ce{H} energy as its reference (Table~\ref{tab:hchain-size-consistency}).  The table
reports total energy per atom.  We define the residual as
$\Delta_N/N=(E_N-N E_1)/N$.  The
phase-including arm remains within $0.22$ mHa per atom of its \ce{H} reference from
\ce{H2} through \ce{H8}.  The completed no-factor trace for \ce{H8}
is numerically valid but lies in a high-energy optimization basin, with
a residual of approximately $22.24$ mHa per atom.  The comparison
therefore shows that the prescribed factor improves both size consistency and training
stability for finite-field NNQMC.
\begin{table}[t]
\centering
\small
\setlength{\tabcolsep}{7pt}
\caption[size consistency test]{\textbf{Size consistency test.}  
The table reports total energies per atom in Ha, for hydrogen chains of bond length $R=10$ bohr
under the magnetic field $\mathbf{B}=(1,0,0) \au$ with and
without the gauge-including factor. The calculations of \ce{H2} to \ce{H8} use a slow field-ramp protocol in which
$B_x$ is increased from zero to $1\ \au$ over the first $60{,}000$ training
iterations, followed by fixed-field optimization to $200{,}000$ iterations.
The single-atom \ce{H} calculations are trained directly at the target field.}
\label{tab:hchain-size-consistency}
\begin{tabular}{lcc}
\toprule
System & Without factor & With factor \\
\midrule
\ce{H} & $-0.3308$ & $-0.3308$ \\
\ce{H2} & $-0.3301$ & $-0.3308$ \\
\ce{H4} & $-0.3253$ & $-0.3310$ \\
\ce{H6} & $-0.3179$ & $-0.3310$ \\
\ce{H8} & $-0.3086$ & $-0.3307$ \\
\bottomrule
\end{tabular}
\end{table}

\begin{figure}[t]
  \centering
  \includegraphics[width=\linewidth]{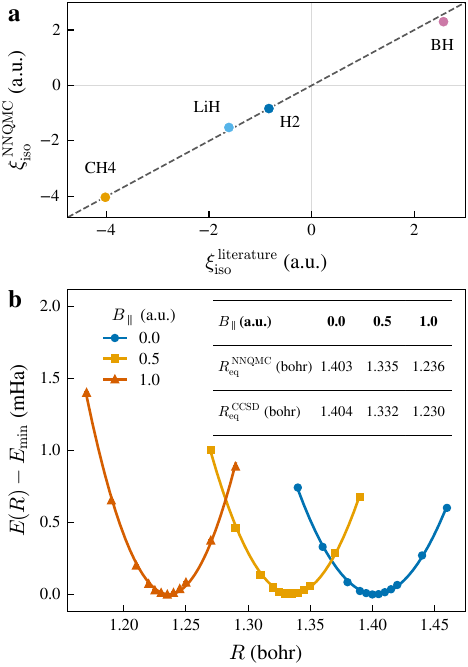}
  \caption[Benchmarking magnetic response and strong-field H2]{\textbf{Benchmarking
  magnetic response and strong-field \ce{H2}.} \textbf{a,} Isotropic magnetizabilities
  from NNQMC compared with the stated literature values.  The plotted literature values
  are fixed-geometry FCI/GIAO for \ce{H2}, an all-electron CCSD(T)/aug-cc-pCV[TQ]Z
  extrapolation for \ce{LiH}, fixed-geometry CCSD(T) for \ce{CH4}, and MCSCF-GIAO for
  \ce{BH}~\cite{ruud1995}.  \textbf{b,} Parallel-field \ce{H2} singlet
  potential-energy curves, shown relative to each fitted minimum.  Markers are NNQMC
  values, curves are guides from local fits, and the inset lists equilibrium distances
  with finite-field CCSD/cc-pVTZ PEC-derived reference values.}
  \label{fig:benchmarks}
\end{figure}

We then benchmark the gauge-including finite-field ansatz in both weak and strong
magnetic fields.  Across the weak-field magnetizability comparisons in
Fig.~\ref{fig:benchmarks}a, NNQMC tracks the corresponding literature values.  For
fixed-geometry \ce{H2} at $R=1.4$ bohr, the finite-field fit gives
$\xi_{\mathrm{iso}}=-0.8318\au$, close to the FCI/GIAO literature values $-0.8313$ and
$-0.8325\au$~\cite{reimann2019,ruud1996h2}.  The parity comparison also includes the
all-electron CCSD(T)/aug-cc-pCV[TQ]Z extrapolated literature value for \ce{LiH} and the
MCSCF-GIAO literature value for \ce{BH}~\cite{lutnaes2009,ruud1995}.

In a parallel strong field, refined \ce{H2} potential-energy scans resolve the
field-induced bond contraction (Fig.~\ref{fig:benchmarks}b).  Quadratic fits give
equilibrium bond lengths of $1.403$, $1.335$, and $1.236$ bohr at $B=0$, $0.5$, and
$1\au$, respectively, compared with $1.404$, $1.332$, and $1.230$ bohr from finite-field
CCSD/GIAO cc-pVTZ potential-energy curves~\cite{zalialiutdinov2025}.  Additional field
points and the perpendicular-field triplet case are reported in the
SI~\cite{suppinfo,lange2012}. Together, these benchmarks reproduce quantitative
weak-field response and strong-field
\ce{H2} bond contraction, strengthening the validity of our framework within a wide
range of field strengths.

\subsection{Application to molecular spectra of white dwarfs}

Strong magnetic fields in white dwarfs can substantially modify molecular
electronic structure. Conversely, molecular spectra may retain a direct
signature of the magnetic field, providing a potential spectroscopic method
for measuring field strengths in white-dwarf atmospheres. Molecules such as
\ce{CN} and \ce{C2} have been identified in carbon-rich DQ white dwarfs
~\cite{venneskawka2025,kowalski2010}. Here, we use our finite-field NNQMC
framework to calculate representative band systems of these molecules, focusing
on the \ce{CN} red system and the \ce{C2} Swan system. We then assess their
sensitivity to white-dwarf magnetic fields.

We calculate matched, field-dependent transition energies for selected
\ce{CN} and \ce{C2} states with the molecular axis parallel to the field.
Excited states are obtained using a symmetry-resolved optimization procedure
that targets the lowest state within a specified symmetry sector (see the
SI~\cite{suppinfo}). For each transition, we define the
corresponding band energy and field-induced shift as
\begin{equation}
  \begin{aligned}
  \mathrm{band}(B)={}&E_{\mathrm{upper}}(B)-E_{\mathrm{lower}}(B),\\
  \mathrm{shift}(B)={}&\mathrm{band}(B)-\mathrm{band}(0).
  \end{aligned}
  \label{eq:band-shift}
\end{equation}
Here, $E_{\mathrm{upper}}$ and $E_{\mathrm{lower}}$ denote the energies of the
selected upper and lower states, respectively. 

\begin{figure}[t]
  \centering
  \includegraphics[width=\linewidth]{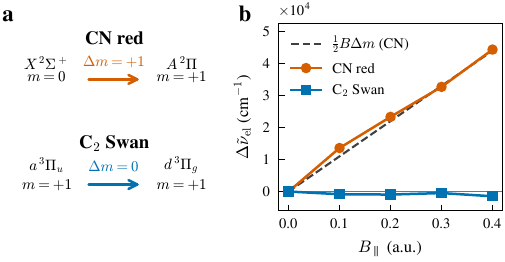}
  \caption[Field-induced electronic shifts of CN and C2]{\textbf{Field-induced
  \ce{CN} red and \ce{C2} Swan transitions.}
  \textbf{a,} Selected axial $m$ sectors for a molecular axis parallel to the applied
  field. \textbf{b,} QMC field-induced shifts defined by Eq.~\eqref{eq:band-shift};
  error bars show propagated within-trace standard errors of the mean. The dashed line
  is the \ce{CN} reference $\tfrac{1}{2}B\Delta m$; the inset resolves the smaller
  \ce{C2} variation.}
  \label{fig:applications}
\end{figure}

The two molecular systems exhibit markedly different electronic responses to the
same external fields. For the \ce{CN} red system,
$A^2\Pi\leftarrow X^2\Sigma^+$~\cite{cnred}, the matched QMC scan reaches a
shift of $+44238\pm136\,\cmone$ at $B=0.4\au$ (
Fig.~\ref{fig:applications}). This large displacement indicates that the
\ce{CN} red system could provide a sensitive spectroscopic diagnostic of
white-dwarf magnetic fields. By contrast, the selected \ce{C2} Swan
transition, $d^3\Pi_g\leftarrow a^3\Pi_u$~\cite{c2swan}, shows a much weaker
response, with shifts not exceeding $1475.5\,\cmone$ in magnitude across the
field range. The sign inconsistency also supports the view that the observed
$100\text{--}300\,\AA$ blueshifts of \ce{C2} Swan bands in DQ white dwarfs
arise primarily from field-independent effects, such as high-pressure helium
~\cite{Hall_2008,kowalski2010}.

This contrast is consistent with the orbital-Zeeman effect. The \ce{CN} red
transition changes the projection of orbital angular momentum from $m=0$ to
$m=+1$, giving $\Delta m=+1$. The leading parallel-field orbital contribution,
$\tfrac{1}{2}B\Delta m$, is therefore nonzero. In the selected \ce{C2} Swan
transition, both states have $m=+1$, so that $\Delta m=0$ and the leading
orbital contribution cancels. This difference in orbital
angular-momentum projection accounts for the pronounced field dependence of
the \ce{CN} red system and the much weaker response of the \ce{C2} Swan 
transition. At field strengths characteristic of magnetic white dwarfs, the
resulting shift of the \ce{CN} red-system transition may therefore provide a
useful probe of the local magnetic field~\cite{ferrario2015}.

\section{Discussion}

We have developed and tested a real-space NNQMC framework for 
molecular electronic structure in external magnetic fields. This 
construction preserves gauge and magnetic-translation covariance while 
reducing the rapid phase oscillation that must be learned variationally. The 
\ce{CH4} translation tests and the separated \ce{H}--\ce{H8} chain calculations 
support the resulting covariance and approximate size consistency.

The benchmark calculations further show that the framework describes  
weak-field magnetic response and strong-field structural changes. Calculated 
magnetizabilities agree with established finite-field values, while the 
parallel-field \ce{H2} potential-energy curves reproduce the expected 
field-induced bond contraction. These results establish the method as a 
consistent approach for finite-field molecular calculations over both 
perturbative and nonperturbative regimes.

Applications to white-dwarf molecular spectra show that different molecular 
transitions can provide complementary responses to the local magnetic field. 
The \ce{CN} red transition is strongly field-sensitive, whereas the selected 
\ce{C2} Swan transition is comparatively insensitive over the same field 
range. This contrast is consistent with their different axial orbital angular momentum: 
the selected \ce{CN} transition changes from $m=0$ to $m=+1$, whereas both 
selected \ce{C2} states have $m=+1$. These results indicate that 
state-resolved molecular spectra, particularly field-sensitive \ce{CN} bands, 
could provide spectroscopic diagnostics of magnetic fields in white-dwarfs. 

\section{Methods}

\subsection{Hamiltonian and local-energy estimator}

All calculations use clamped nuclei and atomic units.  In a chosen symmetric gauge, the
nonrelativistic electronic Hamiltonian is
\begin{equation}
  \begin{aligned}
  \hat H &= \sum_i\frac{1}{2}
  \left[-\iu\nabla_i+\Avec(\rvec_i)\right]^2+V_{\rm Coul}+V_{\rm Zeeman}, \\
  V_{\rm Zeeman} &= g_e\mu_B\Bvec\cdot\hat{\bm S}, \qquad
  \Avec(\rvec) = \frac{1}{2}\Bvec\times\rvec.
  \end{aligned}
  \label{eq:main-hamiltonian}
\end{equation}
Here $\Avec(\rvec)$ is the magnetic vector potential, $g_e$ is electron $g$-factor,
$\mu_B$ is Bohr magneton, and $\hat{\bm S}$ is the total electronic spin operator.
$V_{\rm Coul}$ contains the electron--nuclear, electron--electron, and nuclear--nuclear
Coulomb terms.  The orbital part, namely the kinetic term plus $V_{\rm Coul}$, enters
the magnetic local-energy estimator.  For a fixed spin-projection branch, $V_{\rm
Zeeman}$ is an analytic additive constant. For a complex trial state $\Psi_\theta$, the
local kinetic energy is evaluated as
\begin{equation}
 \begin{aligned}
 E^{\mathrm{loc}}_{\mathrm{kin},i}={}&
 -\frac12\left[\nabla_i^2\log\Psi_\theta+ \left(\nabla_i\log\Psi_\theta\right)^2\right]
 \\
 &{}-\iu\Avec(\rvec_i)\cdot\nabla_i\log\Psi_\theta
 {}+\frac12\Avec(\rvec_i)^2 .
 \end{aligned}
 \label{eq:local-energy}
\end{equation}
The trial state is given in Eq.~\eqref{eq:main-trial-state}.

\subsection{Magnetic-translation covariance}

For a rigid translation, let $\rvec_i'=\rvec_i+\bm t$ and $\bm R_I'=\bm R_I+\bm t$ while
the gauge origin is fixed.  Defining $\bm a_t=\tfrac12\Bvec\times\bm t$ and $F_t(\bm
r)=\sum_i\bm a_t\cdot\rvec_i$, the relations $\Avec(\rvec_i')=\Avec(\rvec_i)+\bm a_t$
and $V_{\bm R+\bm t}(\bm r+\bm t)=V_{\bm R}(\bm r)$ give the magnetic-translation
covariance condition
\begin{equation}
 \Psi_{\bm R+\bm t}(\bm r+\bm t)= \eu^{-\iu F_t(\bm r)}\eu^{\iu\gamma_t}\Psi_{\bm R}(\bm
 r),
 \label{eq:main-magnetic-translation}
\end{equation}
where $\gamma_t$ is coordinate independent.  The weights depend only on
electron--nuclear separations, so the translated geometry has $\bar{\bm R}_i'=\bar{\bm
R}_i+\bm t$ and $\bm C_Z'=\bm C_Z+\bm t$.  Direct substitution into
Eq.~\eqref{eq:main-prescribed-phase} therefore gives
\begin{align}
 \chi_{\rm GI}'(\bm r+\bm t)-\chi_{\rm GI}(\bm r)
 &=-\frac12\sum_i\Bvec\cdot
 \left[\bm t\times(\rvec_i-\bm C_Z)\right] \nonumber\\
 &=-F_t(\bm r)+N\bm a_t\cdot\bm C_Z .
 \label{eq:main-prescribed-translation}
\end{align}
Thus the prescribed factor supplies exactly the coordinate-dependent phase in
Eq.~\eqref{eq:main-magnetic-translation}; the neural factor need transform only by a
coordinate-independent phase.

\subsection{Approximate size consistency}

The same construction is compatible with approximate size consistency in the
dominant-configuration limit.  For two separated neutral fragments $A$ and $B$, let
$N_F$ and $\bm C_F$ be the electron number and nuclear charge centre of fragment $F$,
and define $\bm C_{Z,AB}=(N_A\bm C_A+N_B\bm C_B)/(N_A+N_B)$ and $\bm d_F=\bm C_F-\bm
C_{Z,AB}$.  In the hard-assignment limit, we consider dominant configurations with
chemically faithful assignments, $\sum_{i\in F}\bar{\bm R}_i=N_F\bm C_F$.  Direct
subtraction then gives
\begin{equation}
 \chi_{AB}-\chi_A-\chi_B \simeq-\frac12\Bvec\cdot\sum_{F\in\{A,B\}} N_F\bm C_F\times\bm
 d_F=0,
 \label{eq:main-size-phase-additivity}
\end{equation}
because $\sum_F N_F\bm d_F=0$ and $\bm d_F\times\bm d_F=\bm 0$.  Hence
$\eu^{\iu\chi_{AB}}\simeq\eu^{\iu\chi_A}\eu^{\iu\chi_B}$.  If the separated physical
state factorizes, the neural residual can likewise satisfy $\Phi_{AB}^{\mathbb
C}\simeq\mathcal A[\Phi_A\Phi_B]$, where $\mathcal A$ denotes
fermionic antisymmetrization.  With the interfragment Coulomb, exchange, and overlap
contributions vanishing at large separation, this gives $E_{AB}\simeq E_A+E_B$. Full
derivations and numerical tests of the covariance relations and separated-fragment
argument are provided in the SI~\cite{suppinfo}.

\subsection{Numerical protocols and state-resolved properties}

Numerical settings, uncertainty definitions, projection formulae, and source tables are
provided in the SI~\cite{suppinfo}.  The \ce{CH4} phase-ablation panel uses the
directional $B\parallel z$ fit.

Weak-field magnetizabilities are obtained from even finite-field fits,
$E(B)=E_0+c_2B^2+\cdots$, with $\xi=-2c_2$.  Strong-field curves are Born--Oppenheimer
electronic potential-energy scans at the stated field orientation.  For state-resolved
applications, spin, parity where applicable, and the molecular-axis orbital label
$\Lambda$ are imposed by symmetry restrictions and projection operations that commute
with the finite-field Hamiltonian.

For each targeted transition, lower and upper states use matched field directions and
symmetry prescriptions.  We then form Eq.~\eqref{eq:band-shift}.

% \section*{Author contributions}

\section*{Competing interests}
The authors declare no competing interests.

\bibliography{refs}

@misc{suppinfo,
  author = {L{\"u}, Chengye and Fu, Weizhong and Gong, Xin-gao and Xiang, Hongjun},
  title = {Supplementary Information for {Gauge-including neural-network quantum Monte Carlo for molecules in magnetic fields}},
  year = {2026},
  note = {Available with this manuscript}
}

@article{london,
  author = {London, Fritz},
  title = {Th{\'e}orie quantique des courants interatomiques dans les combinaisons aromatiques},
  journal = {J. Phys. Radium},
  volume = {8},
  pages = {397--409},
  year = {1937}
}

@article{ditchfield1974,
  author = {Ditchfield, Robert},
  title = {Self-consistent perturbation theory of diamagnetism: I. A gauge-invariant {LCAO} method for {NMR} chemical shifts},
  journal = {Molecular Physics},
  volume = {27},
  number = {4},
  pages = {789--807},
  year = {1974},
  doi = {10.1080/00268977400100711}
}

@article{tellgren2008,
  author = {Tellgren, Erik I. and Soncini, Alessandro and Helgaker, Trygve},
  title = {Nonperturbative ab initio calculations in strong magnetic fields using {London} orbitals},
  journal = {J. Chem. Phys.},
  volume = {129},
  pages = {154114},
  year = {2008},
  doi = {10.1063/1.2996525}
}

@article{stopkowicz2015,
  author = {Stopkowicz, Stella and Gauss, J{\"u}rgen and Lange, Kai K. and Tellgren, Erik I. and Helgaker, Trygve},
  title = {Coupled-cluster theory for atoms and molecules in strong magnetic fields},
  journal = {J. Chem. Phys.},
  volume = {143},
  pages = {074110},
  year = {2015},
  doi = {10.1063/1.4928056}
}

@article{bartlett2007,
  author = {Bartlett, Rodney J. and Musia{\l}, Monika},
  title = {Coupled-cluster theory in quantum chemistry},
  journal = {Rev. Mod. Phys.},
  volume = {79},
  number = {1},
  pages = {291--352},
  year = {2007},
  doi = {10.1103/RevModPhys.79.291}
}

@article{lehtola2020diatomic,
  author = {Lehtola, Susi and Dimitrova, Maria and Sundholm, Dage},
  title = {Fully numerical electronic structure calculations on diatomic molecules in weak to strong magnetic fields},
  journal = {Molecular Physics},
  volume = {118},
  number = {2},
  pages = {e1597989},
  year = {2020},
  doi = {10.1080/00268976.2019.1597989}
}

@article{ferminet,
  author = {Pfau, David and Spencer, James S. and Matthews, Alexander G. D. G. and Foulkes, W. M. C.},
  title = {Ab initio solution of the many-electron {Schr\"odinger} equation with deep neural networks},
  journal = {Phys. Rev. Research},
  volume = {2},
  pages = {033429},
  year = {2020},
  doi = {10.1103/PhysRevResearch.2.033429}
}

@article{li2024polarization,
  author = {Li, Xiang and Qian, Yubing and Chen, Ji},
  title = {Electric Polarization from a Many-Body Neural Network Ansatz},
  journal = {Phys. Rev. Lett.},
  volume = {132},
  number = {17},
  pages = {176401},
  year = {2024},
  doi = {10.1103/PhysRevLett.132.176401}
}

@article{deepqmc,
  author = {Hermann, Jan and Sch{\"a}tzle, Zeno and No{\'e}, Frank},
  title = {Deep-neural-network solution of the electronic {Schr\"odinger} equation},
  journal = {Nat. Chem.},
  volume = {12},
  pages = {891--897},
  year = {2020},
  doi = {10.1038/s41557-020-0544-y}
}

@article{nndmc,
  author = {Ren, Weiluo and Fu, Weizhong and Wu, Xiaojie and Chen, Ji},
  title = {Towards the ground state of molecules via diffusion {Monte Carlo} on neural networks},
  journal = {Nat. Commun.},
  volume = {14},
  pages = {1860},
  year = {2023},
  doi = {10.1038/s41467-023-37609-3}
}

@article{lange2012,
  author = {Lange, Kai K. and Tellgren, E. I. and Hoffmann, M. R. and Helgaker, T.},
  title = {A paramagnetic bonding mechanism for diatomics in strong magnetic fields},
  journal = {Science},
  volume = {337},
  number = {6092},
  pages = {327--331},
  year = {2012},
  doi = {10.1126/science.1219703}
}

@article{reimann2019,
  author = {Reimann, Sarah and Borgoo, Alex and Austad, Jon and Tellgren, Erik I. and Teale, Andrew M. and Helgaker, Trygve and Stopkowicz, Stella},
  title = {Kohn--Sham energy decomposition for molecules in a magnetic field},
  journal = {Molecular Physics},
  volume = {117},
  number = {3},
  pages = {277--290},
  year = {2019},
  doi = {10.1080/00268976.2018.1495849}
}

@article{ruud1996h2,
  author = {Ruud, Kenneth and {\AA}strand, Per-Olof and Helgaker, Trygve and Mikkelsen, Kurt V.},
  title = {Full {CI} calculations of the magnetizability and rotational {g} factor of the hydrogen molecule},
  journal = {J. Mol. Struct. THEOCHEM},
  volume = {388},
  pages = {231--235},
  year = {1996},
  doi = {10.1016/S0166-1280(96)80036-4}
}

@article{lutnaes2009,
  author = {Lutn{\ae}s, Ola B. and Teale, Andrew M. and Helgaker, Trygve and Tozer, David J. and Ruud, Kenneth and Gauss, J{\"u}rgen},
  title = {Benchmarking density-functional-theory calculations of rotational g tensors and magnetizabilities using accurate coupled-cluster calculations},
  journal = {J. Chem. Phys.},
  volume = {131},
  number = {14},
  pages = {144104},
  year = {2009},
  doi = {10.1063/1.3242081}
}

@article{zalialiutdinov2025,
  author = {Zalialiutdinov, T. and Solovyev, D.},
  title = {Exploring the properties of light diatomic molecules in strong magnetic fields},
  journal = {arXiv preprint arXiv:2503.17367v2},
  year = {2025},
}

@article{ruud1995,
  author = {Ruud, Kenneth and Helgaker, Trygve and Bak, Keld L. and J{\o}rgensen, Poul and Olsen, Jeppe},
  title = {Accurate magnetizabilities of the isoelectronic series {BeH}$^-$, {BH}, and {CH}$^+$. {T}he {MCSCF}-{GIAO} approach},
  journal = {Chemical Physics},
  volume = {195},
  number = {1},
  pages = {157--169},
  year = {1995},
  doi = {10.1016/0301-0104(95)00052-P}
}

@article{booth2013,
  author = {Booth, George H. and Gr{\"u}neis, Andreas and Kresse, Georg and Alavi, Ali},
  title = {Towards an exact description of electronic wavefunctions in real solids},
  journal = {Nature},
  volume = {493},
  number = {7432},
  pages = {365--370},
  year = {2013},
  doi = {10.1038/nature11770}
}

@article{c2swan,
  author = {Brooke, James S. A. and Bernath, Peter F. and Schmidt, Timothy W. and Bacskay, George B.},
  title = {Line strengths and updated molecular constants for the {C$_2$} {Swan} system},
  journal = {J. Quant. Spectrosc. Radiat. Transfer},
  volume = {124},
  pages = {11--20},
  year = {2013},
  doi = {10.1016/j.jqsrt.2013.02.025}
}

@article{kowalski2010,
  author = {Kowalski, P. M.},
  title = {The origin of peculiar molecular bands in cool {DQ} white dwarfs},
  journal = {Astron. Astrophys.},
  volume = {519},
  pages = {L8},
  year = {2010},
  doi = {10.1051/0004-6361/201015238}
}

@article{Hall_2008,
  doi = {10.1086/586889},
  url = {https://doi.org/10.1086/586889},
  year = {2008},
  month = {may},
  publisher = {},
  volume = {678},
  number = {2},
  pages = {1292},
  author = {Hall, Patrick B. and Maxwell, Aaron J.},
  title = {C2 in Peculiar DQ White Dwarfs},
  journal = {The Astrophysical Journal}
}

@article{cnred,
  author = {Sneden, Christopher and Lucatello, Sara and Ram, R. S. and Brooke, James S. A. and Bernath, Peter},
  title = {Line lists for the {A$^2\Pi$--X$^2\Sigma^+$} (red) and {B$^2\Sigma^+$--X$^2\Sigma^+$} (violet) systems of {CN}},
  journal = {Astrophys. J. Suppl. Ser.},
  volume = {214},
  number = {2},
  pages = {26},
  year = {2014},
  doi = {10.1088/0067-0049/214/2/26}
}

@article{venneskawka2025,
  author = {Vennes, St{\'e}phane and Kawka, Adela},
  title = {The total mass of the close, double degenerate ({DA}+{DQ}) system {NLTT} 16249},
  journal = {Mon. Not. R. Astron. Soc.},
  volume = {536},
  number = {2},
  pages = {1180--1187},
  year = {2025},
  doi = {10.1093/mnras/stae2693}
}

@article{ferrario2015,
  author = {Ferrario, Lilia and de Martino, Domitilla and G{\"a}nsicke, Boris T.},
  title = {Magnetic white dwarfs},
  journal = {Space Sci. Rev.},
  volume = {191},
  number = {1--4},
  pages = {111--169},
  year = {2015},
  doi = {10.1007/s11214-015-0152-0}
}

@article{lai2001,
  author = {Lai, Dong},
  title = {Matter in strong magnetic fields},
  journal = {Rev. Mod. Phys.},
  volume = {73},
  number = {3},
  pages = {629--662},
  year = {2001},
  doi = {10.1103/RevModPhys.73.629}
}

@article{harding2006,
  author = {Harding, Alice K. and Lai, Dong},
  title = {Physics of strongly magnetized neutron stars},
  journal = {Rep. Prog. Phys.},
  volume = {69},
  number = {9},
  pages = {2631--2708},
  year = {2006},
  doi = {10.1088/0034-4885/69/9/R03}
}

@article{ruud1993,
  author = {Ruud, Kenneth and Helgaker, Trygve and Bak, Keld L. and J{\o}rgensen, Poul and Jensen, Hans J{\o}rgen Aa.},
  title = {Hartree--Fock limit magnetizabilities from {London} orbitals},
  journal = {J. Chem. Phys.},
  volume = {99},
  number = {5},
  pages = {3847--3859},
  year = {1993},
  doi = {10.1063/1.466131}
}

@article{stein2016,
  author  = {Stein, Christopher J. and Reiher, Markus},
  title   = {Automated Selection of Active Orbital Spaces},
  journal = {Journal of Chemical Theory and Computation},
  volume  = {12}, number = {4}, pages = {1760--1771}, year = {2016},
  doi     = {10.1021/acs.jctc.6b00156}
}

@inproceedings{psiformer,
  author = {von Glehn, Ingrid and Spencer, James S. and Pfau, David},
  title = {A self-attention ansatz for ab-initio quantum chemistry},
  booktitle = {The Eleventh International Conference on Learning Representations},
  year = {2023},
  eprint = {2211.13672},
  archiveprefix = {arXiv}
}

@article{qian2025deephall,
  author = {Qian, Yubing and Zhao, Tongzhou and Zhang, Jianxiao and Xiang, Tao and Li, Xiang and Chen, Ji},
  title = {Solving and visualizing fractional quantum {Hall} wavefunctions with neural networks},
  journal = {Phys. Rev. Lett.},
  volume = {134},
  pages = {176503},
  year = {2025},
  doi = {10.1103/PhysRevLett.134.176503}
}

@article{teng2025electron_gas,
  author = {Teng, Yi and Dai, Daniel D. and Fu, Liang},
  title = {Solving the fractional quantum {Hall} problem with self-attention neural network},
  journal = {Phys. Rev. B},
  volume = {111},
  number = {20},
  pages = {205117},
  year = {2025},
  doi = {10.1103/PhysRevB.111.205117}
}

@article{abouelkomsan2026fqh,
  author = {Abouelkomsan, Ahmed and Fu, Liang},
  title = {First-Principles {AI} Finds Crystallization of Fractional Quantum {Hall} Liquids},
  journal = {PRX Intelligence},
  volume = {1},
  number = {1},
  pages = {013010},
  year = {2026},
  month = aug,
  doi = {10.1103/9qlr-jp6x}
}

@article{fu2026towards,
  title={Towards stable and accurate electron dynamics via neural network based time-dependent variational Monte Carlo},
  author={Fu, Weizhong and Li, Zhe and Qian, Yubing and Li, Ruichen and Ren, Weiluo and Chen, Ji},
  journal={arXiv preprint arXiv:2606.05850},
  year={2026}
}

@article{pfau2024accurate,
  title={Accurate computation of quantum excited states with neural networks},
  author={Pfau, David and Axelrod, Simon and Sutterud, Halvard and von Glehn, Ingrid and Spencer, James S.},
  journal={Science},
  volume={385},
  number={6711},
  pages={eadn0137},
  year={2024},
  doi={10.1126/science.adn0137},
  publisher={American Association for the Advancement of Science}
}

@article{fu2024variance,
  title={Variance extrapolation method for neural-network variational Monte Carlo},
  author={Fu, Weizhong and Ren, Weiluo and Chen, Ji},
  journal={Machine Learning: Science and Technology},
  volume={5},
  number={1},
  pages={015016},
  year={2024},
  publisher={IOP Publishing}
}

@article{li2022ab,
  title={Ab initio calculation of real solids via neural network ansatz},
  author={Li, Xiang and Li, Zhe and Chen, Ji},
  journal={Nature Communications},
  volume={13},
  number={1},
  pages={7895},
  year={2022},
  publisher={Nature Publishing Group UK London}
}

@article{li2024spin,
  title={Spin-symmetry-enforced solution of the many-body Schr{\"o}dinger equation with a deep neural network},
  author={Li, Zhe and Lu, Zixiang and Li, Ruichen and Wen, Xuelan and Li, Xiang and Wang, Liwei and Chen, Ji and Ren, Weiluo},
  journal={Nature Computational Science},
  volume={4},
  number={12},
  pages={910--919},
  year={2024},
  publisher={Nature Publishing Group US New York}
}

@article{li2024computational,
  title={A computational framework for neural network-based variational Monte Carlo with Forward Laplacian},
  author={Li, Ruichen and Ye, Haotian and Jiang, Du and Wen, Xuelan and Wang, Chuwei and Li, Zhe and Li, Xiang and He, Di and Chen, Ji and Ren, Weiluo and others},
  journal={Nature Machine Intelligence},
  volume={6},
  number={2},
  pages={209--219},
  year={2024},
  publisher={Nature Publishing Group UK London}
}

@article{fu2026empowering,
  title={Empowering neural network-based quantum {Monte Carlo} with local pseudopotentials},
  author={Fu, Weizhong and Fujimaru, Ryunosuke and Li, Ruichen and Liu, Yuzhi and Wen, Xuelan and Li, Xiang and Hongo, Kenta and Wang, Liwei and Ichibha, Tom and Maezono, Ryo and others},
  journal={Nature Computational Science},
  year={2026},
  month=jul,
  doi={10.1038/s43588-026-01008-7}
}

@article{scherbela2025accurate,
  title={Accurate ab-initio neural-network solutions to large-scale electronic structure problems},
  author={Scherbela, Michael and Gao, Nicholas and Grohs, Philipp and G{\"u}nnemann, Stephan},
  journal={arXiv preprint arXiv:2504.06087},
  year={2025}
}

@phdthesis{liu2018visualization,
  author = {Liu, Yu},
  title = {The visualization of chemical bonding motifs from many-electron wavefunctions: Dynamic Voronoi Metropolis Sampling},
  school = {UNSW Sydney},
  year = {2018},
  doi = {10.26190/unsworks/3420},
  url = {https://unsworks.unsw.edu.au/handle/1959.4/60209}
}

\end{document}